\documentclass{doublecol}
 
\usepackage{amssymb}
\usepackage{amsbsy,natbib}
\usepackage{amsthm}
\usepackage{amsfonts,amsmath,bm}
\usepackage[dvipsone]{epsfig}
\usepackage{caption2}
\usepackage{graphicx,epsf}
\usepackage{epsf}

\begin{document}

\def\p3m{P${}^3$M}
\def\ap3m{AP${}^3$M}
\def\Hydra{{\small HYDRA} }
\def\eg{{\it e.g.\ }}
\def\etc{{\it etc.\ }}
\def\ie{{\it i.e.\ }}
\def\cf{{\it c.f.\ }}
\def\etal{{\it et al.\thinspace}}
\def\fig{figure}
\def\Fig{Figure}
\def\figs{figures}
\def\Figs{Figures}
\def\url{}

\LRH{A New Fast Parallel Statistical Measurement
Technique for Computational Cosmology}

\RRH{R. J. Thacker and H. M. P. Couchman}

\VOL{1}
 
\ISSUE{1/2/3}
                                                                                
\PUBYEAR{2004}

\setcounter{page}{1}

\BottomCatch

\title{A New Fast Parallel Statistical Measurement 
Technique for Computational Cosmology}

\authorA{Robert J. Thacker*}

\affA{Department of Physics, Engineering Physics and Astronomy,\\
Queen's University, Kingston, Ontario, Canada K7L 3N6\\
E-mail: thacker@astro.queensu.ca \\ *Corresponding author}

\authorB{H. M. P. Couchman}

\affB{Department of Physics and Astronomy\\
McMaster University\\
1280 Main St. West, Hamilton, Ontario, Canada L8S 4M1\\
E-mail: couchman@physics.mcmaster.ca}

\begin{abstract}
Higher order cumulants of point processes, such as skew and kurtosis,
require significant computational effort to calculate. The traditional
counts-in-cells method implicitly requires a large amount of computation
since, for each sampling sphere, a count of particles is necessary.
Although alternative methods based on tree algorithms can reduce execution
time considerably, such methods still suffer from shot noise when
measuring moments on low amplitude signals. We present a novel method for
calculating higher order moments that is based upon first top-hat
filtering the point process data on to a grid. After correcting for the
smoothing process, we are able to sample this grid using an interpolation
technique to calculate the statistics of interest. The filtering technique
also suppresses noise and allows us to calculate skew and kurtosis when
the point process is highly homogeneous. The algorithm can be implemented
efficiently in a shared memory parallel environment provided a data-local
random sampling technique is used. The local sampling technique allows us
to obtain close to optimal speed-up for the sampling process on the
Alphaserver GS320 NUMA architecture.
\end{abstract}

\KEY{Cosmology; statistics; parallel programming; simulation}

\REF{to this paper should be made as follows: Thacker, R.J.
and Couchman, H.M.P. (2006) `A New Fast Parallel Statistical Measurement
Technique for Computational Cosmology'
Int. J. High Performance
Computing and Networking, Vol. 3, Nos. 2, pp.xx--yy.}

\BIO{Robert Thacker is an Adjunct Assistant Professor
of Physics at Queen's University, Kingston, Ontario, Canada.
He received his PhD
degree from the University of Alberta in 1999.
 Since 1995 he has been conducting research in cosmology and is
a specialist in numerical simulation and parallel computing.
\newline
\newline
Hugh M. P. Couchman is a Professor of Physics at McMaster 
University, 
Hamilton, Ontario, Canada. He received his PhD from Cambridge 
University in 1986. Since that time he has pursued research in 
computational cosmology and is co-developer of a number of 
tools that have aided breakthroughs in numerical modelling. 
Since 2004 he has served as the 
Scientific Director 
of the SHARCNET computing collaboration and is recognized as a pioneer 
of a number of 
Canadian HPC initiatives.}

\maketitle

\section{Introduction}
\label{sect:introduction}
\noindent Modern cosmology, the study of the large 
scale 
structure and 
evolution of 
our universe \citep{pea}, has advanced to the point where we can now 
answer some
very fundamental questions about the distribution of matter within our
universe. Ever since Einstein postulated the theory of General Relativity
and, together with De Sitter \citep{pais}, showed how it could be 
applied 
to the 
universe as a whole, generations
of physicists have pondered on the question of what is the overall
geometry of our universe. Within the past few years observations of the 
relic microwave radiation from the ``Big Bang'' \citep{wmap} 
have shown that the 
universe 
exhibits a geometry quite unlike that expected from theoretical prejudices 
alone.

Although on the largest scales the distribution of matter within our
universe is both homogeneous and isotropic, on smaller scales---less 
than 
1/20th the size of
our visible universe---it is highly inhomogeneous. Even though the matter 
distribution of the universe was exceptionally smooth 300,000 years after 
the creation event \citep{KT}, over billions of years the ubiquitous 
attraction of the 
gravitational force amplifies the minute fluctuations in the early matter 
distribution into the structure 
we see today.
Moreover, the current best theories of 
structure formation
suggest that the matter distribution we observe is formed in a 
`hierarchical clustering' manner with the small structures merging to form 
larger ones 
and so forth \citep{pea}. This growth of structure is accelerated by 
an unseen massive `dark matter' component in our universe.  Although dark
matter cannot be observed directly, there is sufficient evidence within
observations to conclusively infer its existence. Modifications to
Newton's equations, to change gravitational accelerations on large 
scales,
have had limited success, and cannot presently be cast in
a form
compatible with General Relativity \citep{stm}. 

Understanding the distribution of matter within our local universe can
tell us much about the cosmic structure formation process. While on the
very largest scales gravity is the dominant force, on smaller scales gas
pressure forces, from the gaseous inter-galactic (IGM) and inter-stellar
mediums (ISM), can play a significant role.  In clusters of galaxies, for
example, hydrodynamic forces produced by the IGM lead to a distribution of
gas that is held close to hydrostatic equilibrium.
Indeed,
understanding the interaction between the ISM and the stars that condense
out of it, is currently one of the hottest
research areas in cosmology \citep{rt01}. Since if we can understand 
this 
process we
are much closer to being able to infer how the galaxies we observe relate
to the underlying distribution of dark matter that dominates the evolution
of structure.

 Although we are yet to absolutely determine the relation 
between galaxies and dark matter, measuring the distribution of 
 galaxies is the only way of infering the 
distribution of all matter (visible or not). Measurements 
of the speed of 
recession of 
local 
galaxies, led \cite{ep} to form the distance-redshift 
relation now know as `Hubble's Law', which has become a bedrock for the 
development of cosmological theory. Although 
modern surveys of galaxies 
use an updated, and more accurate, form of the distance-redshift relation 
to uncover the spatial distribution of galaxies, the principles involved 
remain the same as those used by Hubble.
 
\begin{figure*}[!t]
\includegraphics[scale=0.875]{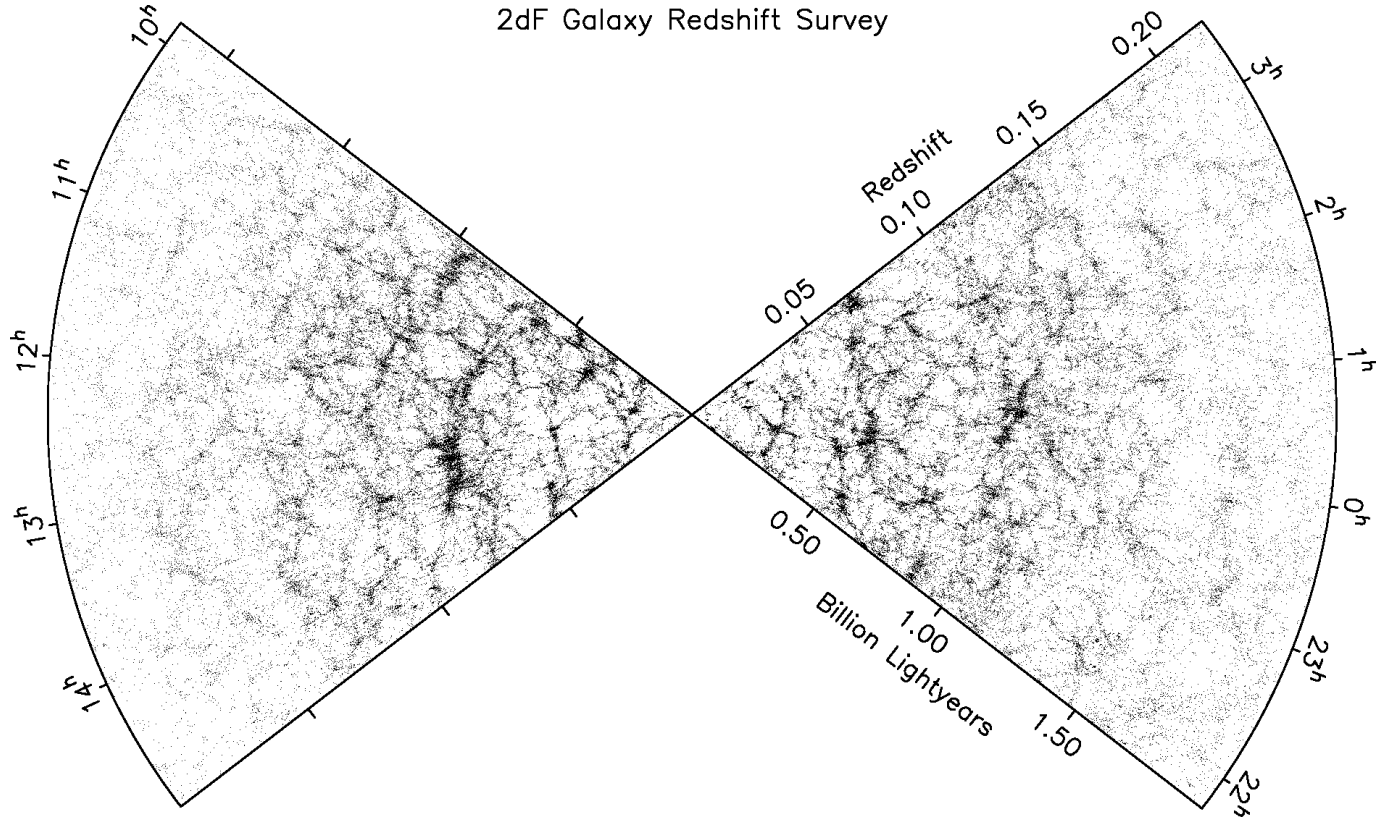}
\caption{Distribution of galaxies in the two main slices from the 2dF 
galaxy redshift survey. Each point represents a galaxy, and they combine 
to trace filament and wall structures in three dimensions. The geometry of 
the distribution is directly related to the statistical properties of the 
conditions in our universe following the ``Big Bang''.}  
\label{2df}
\end{figure*}

Aided by highly automated observing and computer driven data analysis, 
a new generation of high quality galaxy redshift surveys is mapping our
local Universe with exquisite precision. The 2 degree field 
\citep{2df}  and Sloan Digital Sky Survey \citep{SDSS} provide
astronomers with a survey of the local universe out to a redshift of
$z\simeq 0.3$, and contain over 200,000 and one million (when complete)  
redshifts respectively. In figure \ref{2df} we show the distribution of 
galaxies 
for the 2dF survey to give an visual impression of the type of 
inhomogeneity observed.

Traditionally, one of the primary goals of analysis of redshift surveys is
the calculation of the two point auto-correlation function (2-pt CF). The
large sample volumes provided by 2dF and the SDSS have allowed the 2-pt CF
to be calculated with great accuracy.  While the initial conditions
produced by the ``Big Bang'' are widely believed to exhibit Gaussian
statistics (\eg Kolb and Turner, 1990), the formation of structure by 
gravitational
instability introduces non-Gaussian features into the statistics of the 
matter
distribution. Hence, the 2-pt CF cannot be a complete descriptor of the
underlying matter distribution at late times. Astronomers were aware of 
this issue
comparatively early in the development of the field, and the theoretical
basis for calculating higher order statistics was developed through the
1970's (see \cite{P80} for a detailed summary).

Early attempts to measure higher order moments of the mass distribution,
via the counts-in-cells method (again see \cite{P80}), suffered from
inadequate sample size. Because higher order moments tend to be
progressively dominated by the most dense regions in a given sample,
ensuring that adequate sampling has been performed is of utmost
importance. Ensuring low sample variance is also necessary, and given one
sample the only way to check this is to analyse sub-samples, which rapidly
depletes the available information.

From a theoretical perspective, higher order statistics are interesting 
in relation to gravitational perturbation theory and the evolution of 
non-linear gravitational clustering. Analyses examining the accuracy of 
numerical simulation methods often rely upon higher order statistics. This 
is especially important in the study of gravitational clustering in `scale 
free' universes \citep{PC90}. The development of fast, parallel, 
statistical algorithms is vital to progress in this arena. While the 
development of parallel simulation algorithms has advanced forward rapidly 
(\eg Thacker et al., 2003) development of parallel analysis tools has 
lagged behind.
This is partially due to the fact that the benefits of developing a 
parallel analysis code can be shorted lived because the required analyses 
can change rapidly (much faster than the simulation algorithms 
themselves). The rapid development times available on shared memory 
parallel machines make them an ideal complement to large distributed 
memory machines which most simulations are now run on.

Although throughout this paper we discuss the application of our new
method to cosmology, it can be applied equally well to the statistics of
any point process. Indeed the terms `particle' and `point' are often used 
interchangeably. The method can also be modified to apply to different
dimensions, although in 2 dimensions the gains are expected to be less 
significant
due to the reduced amount of work in the counts-in-cells method.

The layout of this paper is as follows: in section \ref{sect:stats}, we
quickly review the statistics we wish to calculate. This is followed by an
explicit description of our new algorithm, and an examination of its
performance.  Next we present a brief case study on applying our algorithm
to cosmology and conclude with a brief summary.

\section{Statistics: Moments and Correlation Functions}
\label{sect:stats}

Due to space limitations a full discussion of the counts-in-cells method,
and how it is related to higher order moments, is beyond the scope of this
paper. However an excellent discussion of counts-in-cells and statistical
measurement processes may be found in \cite{P80}. For completeness, we
briefly summarize the statistics we are interested in measuring.

The 2-pt CF,  $\xi(r)$, measures the radial excess/deficit
over Poisson noise for a point process. It is defined 
in terms of the joint probability, $\delta P$, of finding objects in 
volume elements 
$\delta V_1$ and $\delta V_2$ separated by a radial distance $r_{12}$, 
viz,
\begin{equation}
\delta P=n^2(1+\xi(r_{12}))\delta V_1 \delta V_2,
\end{equation}
where $n$ is the average number density of the point process. The Fourier 
transform pair of the 2-pt CF is the power spectrum, $P(|{\bf k}|)$, 
\begin{equation}
P(|{\bf k}|)= {1 \over (2\pi)^3} \int_V e^{-i {\bf k}.{\bf r}} \xi(|{\bf 
r}|) d^3 
r  
\end{equation}
which is used to describe the statistics of the initial density field 
in cosmology. 

The joint probability idea can be generalized to n-pt processes, for 
example, the reduced 3-pt CF is defined by;
\[
\delta P=n^3\delta V_1 \delta V_2\delta V_3 \times
\]
\begin{equation}
(1+\xi(r_{1})+\xi(r_{2})+\xi(r_{3})+\zeta(r_{1},r_{2},r_{3})),
\end{equation}
where $r_1$,$r_2$ and $r_3$ are defined by the triangle described by the 
three points under consideration. For cosmology, the assumptions of 
homogeneity and isotropy require that $\zeta(r_{1},r_{2},r_{3})$ be a 
symmetric function of these three lengths. Higher order correlation 
functions follow in a logical manner.

Using the counts-in-cells method, it can be shown that the second central 
moment $\mu_2=\left<(N-nV)^2\right>$, where N is the count of points 
within 
spheres of 
radius $r$ (and volume $V$), is given by
\begin{equation}
\mu_2=nV+n^2 \int_V dV_1 dV_2 \xi(r_{12}).
\end{equation}
The third central moment $\mu_3=\left<(N-nV)^3\right>$, is given by
\begin{equation}
\mu_3=3\mu_2 - 2 nV + n^3 \int_V dV_1 dV_2 dV_3 \zeta
\end{equation}
Both these equations show how integrals over the correlation functions 
enter in to calculations of the central moments. Relationships for the 
higher order moments can be constructed, but rapidly become lengthy to 
calculate (\eg Fry and Peebles, 1978).

The final definition we require is one that relates higher order 
cumulants to the variance.
  To aid our discussion we introduce the 
following 
notation: the over-density of a point process relative to the mean 
density, 
$\bar{\rho}$, is given by $\delta({\bf x})=\Delta \rho({\bf 
x})/\bar{\rho}$ 
where $\Delta 
\rho= \rho({\bf x})-\bar{\rho}$ is the local deviation from the average 
density. Although this is most usually recognized as a 
continuum description, it also provides a useful 
construct for our discussion of point processes. For example,
since 
the local density of particles in the counts-in-cells method is given by 
$N/V$, $\delta({\bf x})\simeq (N/V-n)/n$. From this definition of 
$\delta({\bf x})$ the 
$n$-th 
order connected moments of 
the 
point process define the `$S_p$' statistics via the following 
definition\footnote{The $S_p$ statistics are motivated by the assumption 
that, given the 2-pt CF, $\xi(r)= (r_0/r)^\gamma$, the $n$-pt 
correlation functions scale as 
$\xi^{(n)}(\lambda x_1,...,\lambda x_n) = \lambda^{-\gamma(n-1)}  
\xi^{(n)}(x_1,...,x_n)$, see Balian and Schaeffer (1989).}: 

\begin{equation}
\left< \delta^p \right> = S_p \left<\delta^2 \right>^{p-1}. 
\end{equation}

The $S_p$ statistics play a central role in analysis of redshift surveys. 
To date, up to $S_9$ has been calculated by researchers \citep{S96}.

\section{The Smoothed Field Algorithm (SFA)}
\label{sect:tophat}

While the counts-in-cells method is conceptually beautiful in its relation
to the $S_p$ statistics, it is computationally strenuous to calculate. As
the radius of the sampling sphere becomes larger, on average the work to
calculate the count within the sphere will grow at a cubic rate. In
reality the situation can be potentially worse, since inefficiencies in
particle book-keeping can appear (\ie having to search far down
tree-nodes, or equivalently searching through very dense cells in a grid
code). To counter this problem one can use a hierarchical (tree) storage
of counts in cells on a grid, as discussed in \cite{S99}. This greatly
improves calculation time, since the summation over particles within cells
is much reduced at large radii. Using this method it has been reported
that $10^9$ samples from a data set with 47 million particles can be
generated in 8 CPU hours.

The basis of our alternative `smooth field algorithm' is that each
counts-in-cells 
value is a discrete sample of the local density field smoothed over the
scale of the sample sphere. In the continuum limit of an infinite number
of particles, defining the density  $\delta({\bf x})$, the sampled
value $\delta_s ({\bf x})$ can
be written as an integral over the spherical top-hat function $W(r,r_f)$
\begin{equation}
W(r,r_f)=
\begin{cases}
 1,  & \text{$r\leq r_f$}; \\
 0, & \text{$r>r_f$},
\end{cases}
\end{equation}
of radius $r_f$ and the raw density field $\delta({\bf x})$, to give,   
\begin{equation}
\delta_{s}({\bf x})={1 \over V_{TH}} \int_{V} \delta({\bf x}+{\bf r})
W(|{\bf
r}|,r_f)
d^3 r,
\end{equation}
where $V$ is the volume of the periodic sample region and $V_{TH}$ the  
volume of the sample sphere (a 3 dimensional top-hat).
Via the Convolution Theorem, the Fourier
transform
of $\delta_s({\bf x})$, namely, $\hat{\delta}_s({\bf k})$ is given by
\begin{equation}
\hat{\delta}_s({\bf k})=\hat{\delta}({\bf k}) \hat{W}((|{\bf k}|,r_f).
\end{equation}
Thus we can quickly calculate the {\em entire} $\delta_{s}({\bf x})$ field 
by
Fourier methods.

The discrete calculation of counts can be expressed in almost
the same way, except that the continuous density field is replaced by a 
discrete sum of three dimensional Dirac delta functions, $\delta^{D}({\bf 
x})$,
\begin{equation}
N_s({\bf x})={1 \over V_{TH}} \int_{V} \sum_{i=1}^{N_p} \delta^{D}({\bf 
x}+{\bf
r}-{\bf x}_i)   
W(|{\bf
r}|,r_f)
d^3 r
\end{equation}
where $N_p$ is the number of particles in the simulation, and ${\bf x}_i$
gives the position of particle $i$. In the counts-in-cells method
the integral over
the volume is replaced by a summation within the given search volume 
$V_{TH}$.

To connect these two approaches all that is needed is a smoothing function
that will convert a discrete set of points to a continuous density field.
We require a smoothing function, $A({\bf x})$, which can be summed over
the particle positions to reproduce a smooth field $\delta({\bf x})$.   
Provided we can do this, we can use Fourier methods to precalculate all of
the required $\delta_s({\bf x})$ values and greatly reduce the amount of
work. In practice it will be necessary to define a discrete density on a
grid, and then use an interpolation process to provide a continuum limit.
The smoothing idea has been studied in great depth (see \cite{HE81} for
explicit details) and there exists a series of computationally efficient
smoothing strategies that have good Fourier space properties, as well as
having well defined interpolation function pairs. The most common
smoothing function (`assignment function') mechanisms are `CIC'
(Cloud-in-Cell), and `TSC' (Triangular Shaped Cloud). Cloud-in-cell
interpolation provides a continuous piece-wise linear density field, while
TSC has a continuous value and first derivative. The only potential issue
of difficulty is that sampling a continuous periodic variable at discrete
points means that the Fourier domain is finite and periodic and thus has
the possibility of being polluted by aliased information (with images
separated by $2\pi/L$ where L is the size of the period). In practice, the
higher order assignment functions have a sufficiently sharp cut-off in
Fourier space that this is not a significant problem\footnote{See
\cite{HE81} for a discussion of this point. Aliases can only be removed
completely by assigning information to all points on the sampling grid for
each point/particle, which it too computationally expensive to be
feasible.}.

Having established that we can convert our discrete set of points into a
continuous density defined by a grid of values and an interpolation
function, we must decide upon the size of grid to be used. The initial
configuration of points (corresponding to a low amplitude power spectrum)
is such that the majority of neighbouring particles have separations close
to the mean inter-particle separation $N_p^{1/3}$. Therefore, for this
configuration we use a grid defined such that $L^3=N_p$. This is
beneficial on two counts: firstly, the grid requires a comparatively small
amount of memory to store than the particle data, and secondly, it
captures almost all the density information stored in the particle
distribution (since most particles are separated by sizes close to the 
grid spacing).

To  summarize, the
steps in the SFA are as follows:
\begin{enumerate}
\item{Use an assignment function, $A({\bf x})$, to smooth the mass ($m$)  
associated
with each of
the particles on to a grid. This creates the grid
representation of
the density field, $\rho({\bf x})$:
\[
\rho({\bf x})={m \over V}\sum_{i=1}^{N_p} A({\bf x}_i-{\bf x})   
\]}
\item{Fourier transform the density field $\rho({\bf x})$ to form
$\hat{\delta}({\bf
k})$}
\item{Multiply by $G(k)$, the product of the Fourier transform of the
real
space top-hat
filter
($\hat{W}(k,r_f)=3(\sin(kr_f)-kr_f\cos(kr_f))/kr_f^3$) and the
inverse of the assignment
function filter, which includes an alias sum out to two images}
\item{Fourier transform the resulting field back to real space} 
\item{Calculate $\delta_S({\bf x})$ at all sampling positions using the 
interpolation function
pair to the original assignment function $A({\bf x})$}
\item{Calculate desired statistics}

\end{enumerate}

In this paper we have used a 3rd order polynomial assignment function
(`PQS', see Hockney and Eastwood, 1988) which is defined (in 
1-dimension) by;
\begin{equation}
A_1(x)=
\begin{cases}
 {2 \over 3} + |x|^2\left( {|x| \over 2} -1 \right),  & \text{$|x|\leq 
1$}; \\
 {1 \over 6}(2 - |x|)^3,  & \text{$1<|x|\leq 2$}; \\
 0, & \text{$|x|>2$},
\end{cases}
\end{equation}
and the 3-dimensional function is defined $A(x,y,z)=A_1(x)A_1(y)A_1(z)$.
Note that $A(x,y,z)$ is not an isotropic function, which in this case is
beneficial for speed, since it is unnecessary to calculate a square root.
It also simplifies calculating the Fourier transform of the assignment
function since all the dimensions are now separable. Note that $A_1(x)$
has a comparatively wide smoothing profile, and therefore its Fourier
transform is a strongly peaked function with good band-limiting
properties. This is advantageous for dealing with the aliasing problem
mentioned earlier. Indeed, the Fourier transform of $A_1(x)$ is:
\begin{equation}
\hat{A}(k_x)=\left({\sin(k_x/2)\over k_x/2}\right)^4,
\end{equation}
which has a $1/k^4$ suppression of power.
This is sufficiently sharp to ensure that only the first and second images
need be accounted for in $G(k)$ (the Green's function associated with
top-hat filtering and the assignment process).

\section{Performance Comparison}
\label{sect:perf}
\begin{figure}[!t]
\rotatebox{90}{%
\begin{minipage}{2.5in}
\includegraphics[totalheight=3.25in]{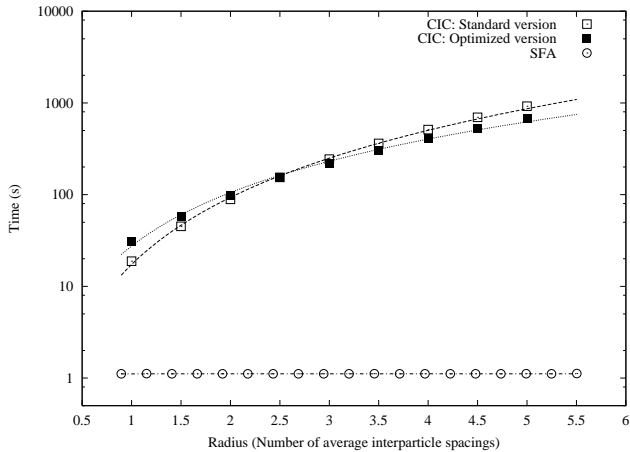}
\end{minipage}
}%
\caption{Comparison of speed for two versions of the counts-in-cells (CIC)
method versus SFA for $2.6\times10^5$ particles and $2.1\times10^6$ sample
points at different smoothing radii. Least squares fits are given for all
data. The first counts-in-cells method is a straight summation over
particles contained within the sampling sphere, while the second method is
optimized to store a count of particles within chaining cells. Provided a
chaining cell lies within the radius of the sampling sphere then the sum
within the chaining cell is not necessary. }
\label{comp} 
\end{figure}

Before proceeding to parallelize the algorithm, it is instructive to
compare the speed of the serial algorithm as compared to the
counts-in-cells method. In figure \ref{comp} we show the time to 
calculate
$2.1\times10^6$ samples on $2.6\times10^5$ points as a function of the
sample radius.  A (logarithmic) least-squares fit showed that the time for
the standard counts-in-cells method (version 1) grows as $r^{2.5}$, which
is slightly lower than the expected value of $r^3$. For the second
counts-in-cells algorithm we developed, which is optimized by storing a
list of counts in the chaining cells used to control particle book-keeping
in the code, the dependence with radius was found to be $r^2$. This is
understood from the perspective that most of the work in each sample has  
already been performed in the summation within chaining cells and that the
work for each sample thus becomes dependent on sorting over the cells at 
the surface of the sample area, which is proportional to $r^2$. However   
comparison of both these methods to the SFA shows they are far slower in 
comparison. Because the entire $\delta_s({\bf x})$ field is precalculated
(modulo the interpolation process to non-grid positions) in the SFA
method, the time to calculate the samples is constant as a function of   
radius, and is exceptionally fast.  Based up the data presented in figure
\ref{comp}, we initially estimated being able to calculate $10^9$ sample 
points 
on
a $512^3$ data set in less than 2 CPU hours, which is over 4x faster than
the results reported for tree-optimized counts-in-cells methods
\citep{S99}. We have recently confirmed this result using our parallel
code, which took 6.5 minutes on 32 processors to calculate $10^9$ samples 
on a $512^3$ particle data set produced for a project being conducted at
the Pittsburgh Supercomputing Center.

\section{Parallelization}
\label{sect:para}
Typically when calculating statistics, the value of the sampling radius
(equivalently the top-hat radius) is varied so that the entire sampling
process
must be
repeated many times. Thus the most obvious method of parallelization is to
create several different grids for each smoothing radius and process them
in parallel. However, available memory considerations may well make this
impractical. Instead, it is better to parallelize each
calculation for each radius. This is non-trivial as
the following algorithmic steps must be
parallelized:
\begin{enumerate}
\item Calculation of Green's function
\item Forward FFT of density grid to $k$-space
\item Multiplication of density grid by Green's function
\item Reverse FFT to real space
\item Sum over sample points
\end{enumerate}
The first four items have all been parallelized previously for our main  
simulation code (see Thacker et al., 1998). The final step, while 
appearing to
be somewhat straightforward, must be approached with care (as we shall   
demonstrate).
The obvious issues which need to addressed are (1) ensuring each
thread has a different random seed for sample positions and (2) that the 
sum reduction of the final values across threads is performed.
In practice, both of these issues can be dealt with in very
straightforward
ways using the OpenMP shared memory programming standard. Sum reductions
can be controlled via the {\tt REDUCTION} primitive while different random
seeds can be set using an array of initial values. Parallelization in this
environment turned out to be straightforward.

\begin{figure}[!t]
\centering
\rotatebox{90}{%
\begin{minipage}{2.5in}
\includegraphics[totalheight=3.25in]{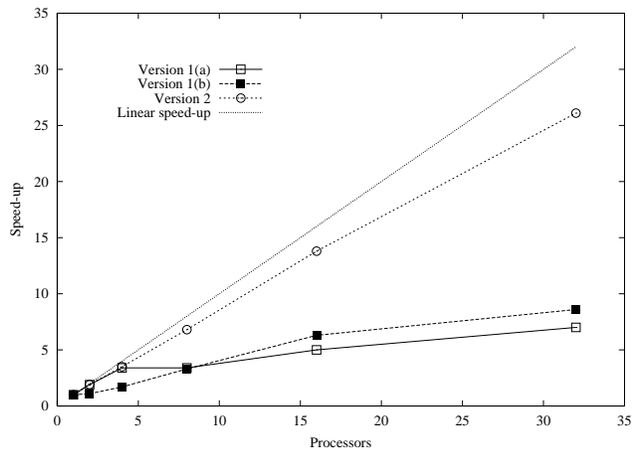}
\end{minipage}
}%
\caption{Comparison of speed-ups for different implementations of the SFA.
Version 1(a) is the standard method with block data decomposition but no
directed assignment within RADs. Version 1(b) forces data and threads to
spread across RADs. Version 2 corresponds to our data local sampling and
is clearly superior. } 
\label{times} 
\end{figure}

Tests on a 32 processor HP GS320 (1 GHz Alpha EV6/7 processors) at the
Canadian Institute for Theoretical Astrophysics (CITA), showed reasonable
speed-up (see figure \ref{times}), 
 but comparatively poor efficiency (22\%) when 32 
processors were used. 
There is also a noticeable step in the
speed-up at 4 to 8 processors. This step is caused by memory for a job
being moved to a second memory domain, or `resource affinity domain' (RAD),
within the machine. The 32 processor machine has 8 RADs in total, connected
via a cross-bar, with 4 processors belonging to each RAD. Latency to remote
RADs is significantly higher than to local RADs, which explains the
increased execution time. Additionally, as the amount of traffic on the
cross-bar between the RADs increases, latencies are known to increase by
very large factors (up to 3000 nanoseconds, Cvetanovic, 2003). This is a
serious bottleneck in the GS320 design which has been removed in the latest
GS1280 machine. Ultimately, to improve performance on the GS320, it is
necessary to increase the locality of the sampling technique to reflect the
locality of memory within the machine, and avoid sending data across the
cross-bar.

\begin{figure*}[!t]
\hspace*{\columnsep}%
  \includegraphics[scale=0.85]{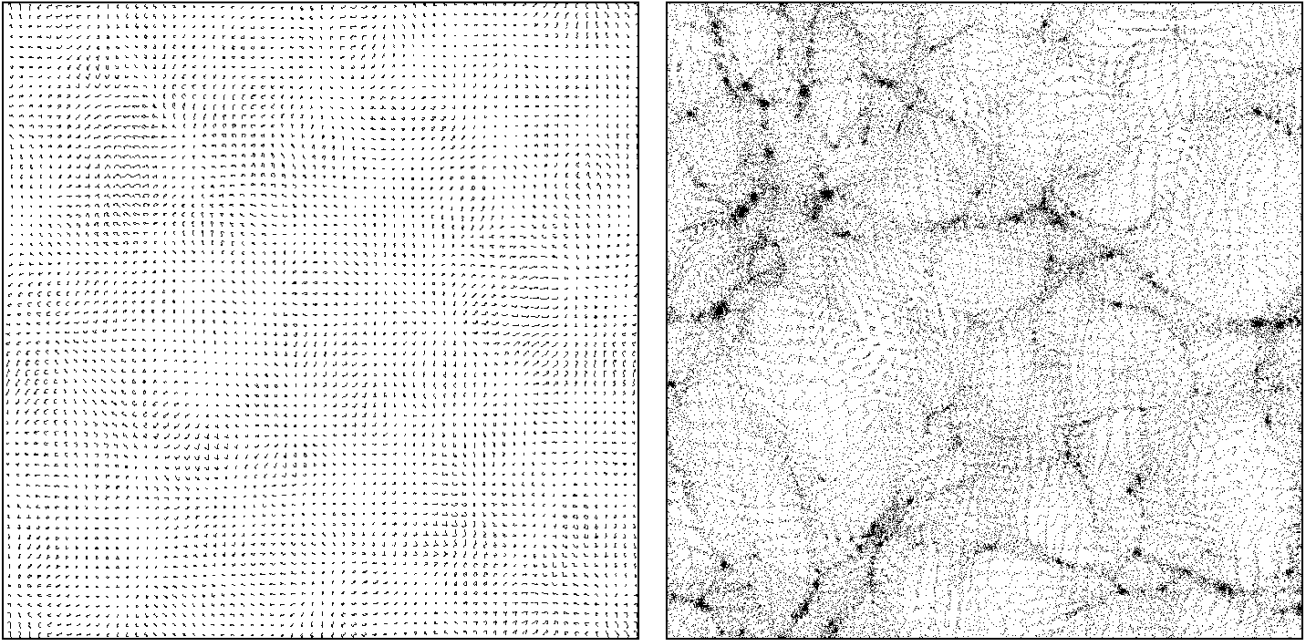}%
\caption{Initial and final point configurations for a slice
through a simulation with
$2.6\times 10^5$ particles.}
\label{dstn}
\end{figure*}

Note that using a block decomposition of data across the RADs
means that locality
is only really necessary in one axis direction.
Therefore, we adopted the following strategy to improve performance:
\begin{enumerate}
\item Block decomposition of the $\delta_s({\bf x})$ grid across
RADs
\item Pre-calculate the list of random positions in the z-axis
\item Parallel sort the list of random positions in increasing z value    
\item Parallelize over the list of z positions, calculating x and y values
randomly
\end{enumerate}
The resulting sample still exhibits Poisson noise statistics and is
therefore valid for our purposes. However, the sample points are now local
in the z direction, which greatly reduces the possibility of remote access
due to the block assignment of data. The scaling improvement for this
method is shown in figure \ref{times}. The improvement is striking. We 
achieved a
1.2x increase in performance for the single processor result alone, while
at 32 processors we have achieved a 4.8$\times$ improvement in 
speed-up and 
a tripling of the  parallel efficiency (82\%). Note that the
speed-up is still not perfect for the improved version. This may be a
bandwidth issue since the interpolation at each sampling point requires 64
grid values, which breaks down into 16 cache lines, with only 8 floating
point calculations performed for all the data in each cache line. Note that
it is unlikely that using the next lowest level of interpolation 
(TSC) would help. TSC requires 27 points grid points per sample,
which is 9 cache lines, with 6 floating point calculations per cache-line.
Thus the overall ratio of calculation to memory fetches is actually
reduced.

\section{Application: Moments in Initial  
Conditions}
\label{sect:sims}
The initial conditions for cosmological structure are prescribed by 
initial density, temperature and velocity fields. Although there is debate
over whether evolution in the early universe (such as magnetic fields) may
induce a non-Gaussian signal in the initial conditions \citep{W94}, most
researchers believe that the density field is Gaussian process, and the
velocity may be derived directly from it. In the absence of non-Gaussian
features, the density field, which is usually discussed in terms of the
linear over-density $\delta$, is completely described by its continuous
power spectrum $P(k)= A k^n$, where A is a normalization constant. This
initially smooth field evolves under gravity to produce the locally
inhomogeneous and biased distribution of galaxies we observe today (see
figure \ref{dstn}, which compare particles positions from initial to final 
outputs).
Early evolution, when $\delta({\bf x})\ll1$, is in the linear regime and
can be described by perturbation theory. As the over-density values
approach and later exceed unity, it is necessary to use simulations to
calculate the non-linear evolution. Thus, ideally, the initial conditions
for simulations should correspond to the latest time that can be followed
accurately by perturbation theory.

\cite{S98} has developed an algorithm for the fast calculation
of the particle positions required for cosmological simulations via 2nd
order Lagrangian perturbation theory (2LPT). We have recently implemented 
this algorithm in parallel using OpenMP. Although 2LPT requires more
computation, it has significant advantages over the standard 1st order
technique (known as the Zel'dovich (1968) approximation) as 
higher
order moments exhibit far less transient deviations at the beginning of
the simulation. Further, one should in principle be able to follow the
initial evolution to slightly later epochs using 2LPT and therefore
begin simulations at a slightly later time. In practice, the transient
deviation issue is most significant.

In general, the more negative the spectral index the faster the initial
transients die away. This is helpful, since most simulations are conducted
with an effective spectral index, $n$, of between -1.5 to -3 (depending on
the size of the simulation volume). Also, although we have focused solely
on particle position statistics in this paper, it is worth noting that a
similar analysis can be applied to velocity fields defined on the point
process. Analysis of the transients in the velocity divergence field,
$\theta=\nabla.v$, shows an even greater improvement when using the 2LPT
method \citep{S98}.

\begin{figure}[!t]
\centering
\centering
\rotatebox{90}{%
\begin{minipage}{2.4in}
\includegraphics[totalheight=3.25in]{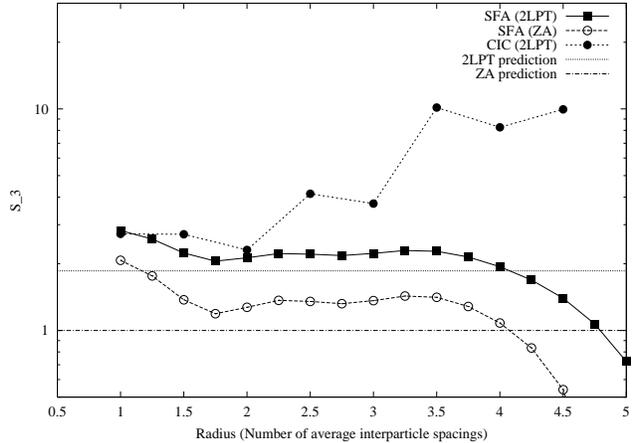}
\end{minipage}
}%
\caption{Comparison of $S_3$ calculated via CIC versus SFA on an n=0
initial condition, with the theoretical result shown for reference.
SFA shows a good match out to 4 inter-particle spacings at which point 
it
begins declining. CIC appears accurate on small scales but rapidly
diverges away from the true signal. We have confirmed that as the
simulation evolves, and the effect of shot
noise is reduced, both methods converge to similar values.}
\label{S3}
\end{figure}

\vspace{0.2cm}
To test whether our new 2LPT code was reproducing the correct results we
have
compared the measured $S_3$ statistics for our 2LPT initial conditions
versus those produces with the Zel'dovich approximation (1st order). At
the
initial expansion factor of $a=1$, the ZA
predicts the following value for $S_3$ \citep{B94};
\begin{equation}
S_3={28 \over 7}-(3+n),
\end{equation}
while 2LPT predicts;
\begin{equation}
S_3={34 \over 7} - (3+n).
\end{equation}
Thus after performing the 2nd order correction the value of $S_3$ should
increase by 6/7.
In figure \ref{S3} we show the calculated values of $S_3$ for two sets of
initial conditions, one created using the ZA and the other with the
additional 2LPT correction. Both the SFA measured values of $S_3$ are high
for this particular set of phases (as compared to the theoretical
prediction), but we have confirmed that alternative
random seeds can produce similar results.  Indeed we have found the
values of $S_3$ are quite dependent upon the phases of the Fourier waves
used, and achieving a value that is asymptotic to the theoretical value  
is extremely difficult. We are currently investigating this phenomenon in
more detail. However, a brief visual inspection of figure \ref{S3} 
provides evidence 
that
the residual, $\Delta$, between the ZA
and 2LPT results is close to $6/7\simeq0.86$.
Analysis of the
set of residuals between the two lines
gives $\Delta=0.88\pm 0.03$ ($1\sigma$ deviation), confirming that
our code is accurately reproducing the difference in $S_3$ values.


\section{Summary and Discussion}
We have presented a new fast algorithm for rapid calculation of one point
cumulants for point processes. Our algorithm is based upon a smoothed
field approach, which reproduces the underlying statistical properties of
the point processes field from which it is derived. The method is
significantly faster \linebreak 
than counts-in-cells methods because the overhead 
of
evaluating the number of particles in a given sphere has been removed.
We are able to
calculate $10^9$ sample points on a $512^3$ data set in less than 2 CPU   
hours, which
is over 4x faster than the results reported for tree-optimized
counts-in-cells methods \citep{S99}. We
also note that while tree methods also lead to very large speed ups, they
are
still subject to noise from the point process for low amplitude signals. 

We are currently applying this new technique to examine the evolution of
high order moments in cosmological density fields at low amplitude levels
and will present our findings elsewhere (Thacker, Couchman and Scoccimarro
in prep). We also anticipate making the codes described in this paper
publically available in the near future.

\section{Acknowledgments}

RJT is partially supported by a CITA National Fellowship.
HMPC acknowledges the support of NSERC and the CIAR. RJT would like to 
thank
Evan Scannapieco and Lars Bildsten for hosting him at U. C. Santa Barbara
where part of this research was conducted. This research utilized
CITA and SHARCNET computing facilities.

\bibliographystyle{apalike}

\end{document}